\documentclass[aps,pre,preprint,superscriptaddress]{revtex4-1}
\usepackage{hyperref}
\usepackage{amsmath}                           
\usepackage{amssymb}                           
\usepackage{dsfont}                            
\usepackage{tabu}
\usepackage{graphicx}
\usepackage{subfigure}

\usepackage{color}

\begin{document}


\title{Collapse of Resilience Patterns in Generalized Lotka-Volterra Dynamics and Beyond.}


\author{Chengyi Tu}
\affiliation{Department of Physics and Astronomy, University of Padua, Via Marzolo 8, 35131 Padova, Italy}

\author{Jacopo Grilli}
\affiliation{Department of Ecology and Evolution, University of Chicago, 1101 E. 57th, Chicago, IL 60637, USA}

\author{Friedrich Schuessler}
\affiliation{Institute of Physics, Albert-Ludwigs-University Freiburg, Hermann-Herder-Straße 3, 79104 Freiburg, Germany}
\affiliation{Network Biology Research Labratories, Technion - Israel Institute of Technology, Haifa 32000, Israel}

\author{Samir Suweis}
\email[]{suweis@pd.infn.it}
\affiliation{Department of Physics and Astronomy, University of Padua, Via Marzolo 8, 35131 Padova, Italy}


\date{\today}

\begin{abstract}
Recently, a theoretical framework aimed at separating the roles of dynamics and topology in multi-dimensional systems has been developed (Gao et al, \textit{Nature}, Vol 530:307 (2016)).
The validity of their method is assumed to hold depending on two main hypothesis: $(i)$ The network determined by the the interaction between pairs of nodes has negligible degree correlations; $(ii)$ The node activities are uniform across nodes on both the drift and pair-wise interaction functions. Moreover, the authors consider only positive (mutualistic) interactions.
Here we show the conditions proposed by Gao and collaborators are neither sufficient nor necessary to guarantee that their method works in general, and validity of their results are not independent of the model chosen within the class of dynamics they considered. Indeed we find that a new condition poses effective limitations to their framework and we provide quantitative predictions of the quality of the one dimensional collapse as a function of the properties of interaction networks and stable dynamics using results from random matrix theory.
We also find that multi-dimensional reduction may work also for interaction matrix with a mixture of positive and negative signs, opening up application of the framework to food-webs, neuronal networks and social/economic interactions.
\end{abstract}

\pacs{}

\maketitle

\section{Introduction}

The fundamental agents of biological or socio-economic systems, from genes in gene-regulatory networks to stock holders in financial markets, act under complex interactions and in general we do not know how to derive their dynamics from first-principle potentials.
In general these interactions are described by pair-wise relations through a matrix (the adjacency matrix) that regulates, typically in non-linear way, the effect of the interactions to the dynamics of the single component.

In particular, there is a rising interest in assessing how interactions determine the stability (or resilience) of dynamical attractors \cite{lyapunov1992general}, i.e. the ability of a system to return after a perturbation to the original equilibrium state \cite{hollnagel2007resilience, rieger2009resilient,walker2004resilience, allesina2012stability}).
Cell biology \cite{huang2005cell, karlebach2008modelling}, ecology \cite{allesina2012stability, suweis2015effect, grilli2017feasibility}, environmental science \cite{drever2006can,barlow2016anthropogenic}, and food security \cite{barthel2013urban, suweis2015resilience} are just some of the many areas of investigation \cite{sheffi2005resilient,folke2006resilience,nelson2007adaptation} where the relation between interaction properties and stability is, although deeply studied, a central open question.
Therefore, understanding the role of system topology in resilience theory for multi-dimensional systems is an important challenge from which our ability to prevent the collapse of ecological and economic systems, as well as to design resilient systems.
Existing methods are only suitable for low-dimensional system \cite{lyapunov1992general}, and, in general, it is not possible to assume that a complex system dynamics can be approximated by one dimensional non-linear equation of the type $\frac{dx}{dt}=f(x,\beta)$, where the ``control'' parameter $\beta$ describes the endogenous effects on the system dynamics.

Recently, Gao et al. \cite{gao2016universal} developed a theoretical framework that collapses the multi-dimensional dynamical behavior onto a one-dimensional effective equation, that in turn can be solved analytically.
They considered a class of equations describing the dynamics of several types (ranging from cellular \cite{alon2006introduction} to ecological \cite{holland2002population, suweis2013emergence} and social systems \cite{pastor2001epidemic}) of multi-dimensional systems with pair-wise interactions.
In this paper, we show under which assumption the proposed method works, we propose new insights on the validity of their framework and we generalize their previous results.
Our work is organized as follows. 
In the next section we summarize the core of Gao et al. framework \cite{gao2016universal}, highlighting the assumption behind their methods. 
In section III we then find that a more general condition poses effective limitations to the validity of the multi-dimensional reduction and we provide quantitative analytical predictions of the quality of the one-dimensional approximation as a function of the properties of the interaction networks and dynamics.
In section IV we then show that the multi-dimensional reduction may work beyond the assumption of strictly mutualistic interactions, thus extending the validity of Gao et al. framework.
We prove analytically our results for generalized Lotka-Volterra and test our conclusions by numerical simulations also for more general dynamics.

\section{Background}

We start by giving a short summary of the multi-dimensional reduction approach for the study of the resilience in complex interacting systems \cite{gao2016universal}.
Gao et al. consider a class of equations describing the dynamics of several types of multi-dimensional systems with two body interactions:
\begin{equation}\label{eq:GenDyn}
\centering
\dot{x}_i = F(x_i) + \sum_{j=1}^S A_{ij} G(x_i, x_j) ,
\end{equation}
where functions $F(x_i)$ and $G(x_i, x_j)$ represent the self-dynamics and interaction dynamics, respectively, and the weight matrix $A_{ij}$ specifies the interaction between nodes.
In particular, they limit their study only to those interaction networks $\mathbf{A}$ that have $(i)$ negligible degree correlations and $(ii)$ all positive entries ($A_{ij}\geq 0$). Moreover, $(iii)$ they assume
that the node activities are uniform across nodes on both the drift and pair-wise interaction functions. 

The resilience of a given fixed point $\mathbf{x}^*$ of a system driven by dynamics Eq. (\ref{eq:GenDyn}) is given by the maximum real eigenvalue $\lambda_1$ of the Jacobian matrix characterizing the linearized dynamics around the fixed point, i.e. $J_{ik}=\frac{\dot{\delta x_i}}{\delta x_k}$.

Gao et al. characterize the effective state of the system using the average nearest-neighbor activity (see Appendix \ref{app:EffFunDer})
\begin{equation}\label{eq:xEffDef}
\centering
x_{eff} = \frac{\sum_{ij} A_{ij} x_j}{\sum_{ij} A_{ij}},
\end{equation}
and an effective control parameter $\beta_{eff}$ that depends on the whole network topology
\begin{equation}\label{eq:betaEffDef}
\centering
\beta_{eff} = \frac{ \sum_{ij} A_{ij} A_{ji} }{ \sum_{ij} A_{ij} },
\end{equation}
i.e., $\beta_{eff}$ is the average over the product of the outgoing and incoming degrees of all nodes.

Finally, they propose that the dynamics of $x_{eff}$ following Eq. (\ref{eq:GenDyn}) can be mapped, independently on $F(x_i)$ and $G(x_i,x_j)$, to the following one-dimensional effective equation:
\begin{equation}\label{eq:GenDynEff}
\centering
\dot{x} = f(\beta,x) = F(x) + \beta G(x, x) ,
\end{equation}
where $\beta$ is the control parameter. 

In this work we will show that: $(a)$ the conditions $(i)$-$(iii)$ above are neither sufficient nor necessary to guarantee that the collapse works in general;
$(b)$ The validity of their results is not independent of the model chosen within the class of dynamics they considered, i.e. does depend on $F$ and $G$.
$(c)$ We show that the restriction $A_{ij} \geq 0$ can be omitted.

We highlight that in this framework the system is assumed to be in one of the stable fixed points, $x^*$ , of Eq. (\ref{eq:GenDynEff}) satisfying $ f(\beta,x^*)=0$ and $\partial_x f|_{x=x^*}<0$.
In other words, for the one-dimensional system given by Eq. (\ref{eq:GenDynEff}) we can calculate analytically the resilience function $x(\beta)$ -- uniquely determined by $f(x,\beta)$ -- which represents the possible states of the system as a function of the parameter $\beta$. Therefore, in order to study the stability or the existence of critical transitions in the complex multi-dimensional system given by Eq. (\ref{eq:GenDyn}) one has to simply calculate $\beta_{eff}$ from the network and analyse the corresponding resilience function $x(\beta)$ corresponding to Eq. (\ref{eq:GenDynEff}).
If the collapse works, then $F(x_{eff}) + \beta_{eff} G(x_{eff}, x_{eff})=0$ is a point on the curve given by $x(\beta)$ (see Figure \ref{fig:Diagram} and Appendix \ref{app:EffFunDer} for mathematical details).
Clearly, this is a powerful result as we can easily study the properties of the one-dimensional non-linear Eq. (\ref{eq:GenDynEff}). 
Therefore our framework is not specific for the theory of Gao et al. (which yields a definite value for $\beta_{eff}$ according to Eq. (\ref{eq:betaEffDef})), but explores the validity of the one-dimensional reduction for any possible value of $\beta$.

\section{Resilience patterns for generalized Lotka-Volterra dynamics}\label{GLVsec}

In order to better understand the relevance of conditions $(i)$ and $(ii)$ on the validity of the results of Gao et al, we consider a simplified setting where both conditions $(i)$ and $(ii)$ are satisfied.
By considering $F(x) = \alpha x$ and $G(x,y) = x y$, the condition $(ii)$ is valid by definition.
In this case the dynamics is defined by the generalized Lotka-Volterra (GLV) equations:
\begin{equation}\label{eq:GLVDyn}
\centering
\dot{x}_i = \alpha x_i + x_i \sum_{j=1}^S A_{ij}x_j ,
\end{equation}
where $\alpha$ is the intrinsic growth rate, and $S$ is the number of species in the community.
The interaction matrix $\mathbf{A}$ is taken to be a random matrix, so that condition $(i)$ is always satisfied.

The advantage of using GLV dynamics is that we have an analytical solution for the stationary state $\mathbf{x^{*} = -\mathbf{A}^{-1} \cdot \boldsymbol{\alpha}}$ as a function of the interaction network $\mathbf{A}$.
Moreover, this solution is globally stable in the positive orthant if $\mathbf{A}$ is negative definite \cite{volterra1936leccons,grilli2017feasibility}.
Finally the corresponding one-dimensional analytical effective equation for GLV dynamics reads as $\frac{dx}{dt} = \alpha x + \beta x^2$, whose feasible ($x(\beta)>0$) and stationary solution is:
\begin{equation}\label{eq:GLVSol}
\centering
x (\beta)= - \alpha/ \beta,
\end{equation}
with $\alpha/ \beta <0$. For values of $\alpha/ \beta>0$, the solution exists, but is not meaningful.

\begin{figure}[!htbp]
	\begin{center}
		\includegraphics[width=0.6\columnwidth]{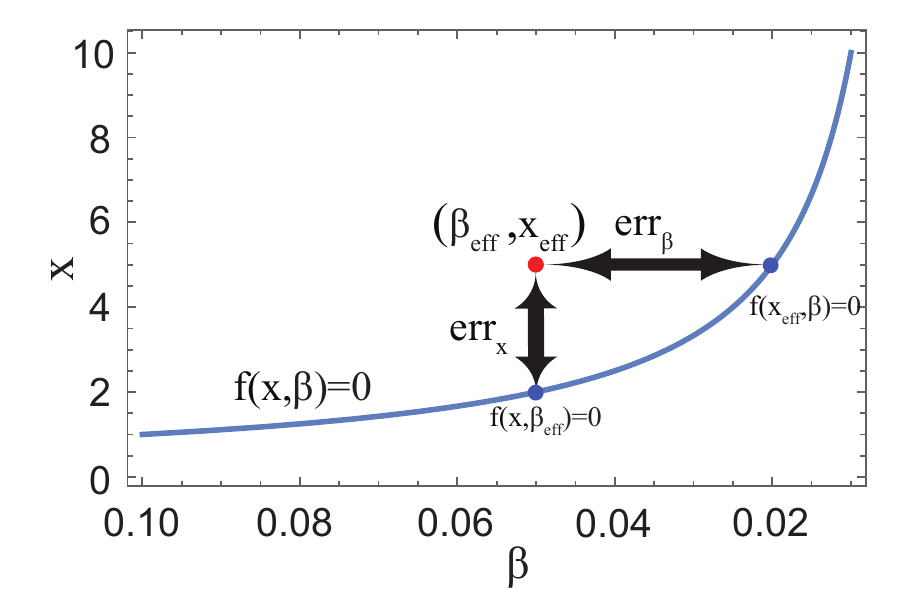}
		\caption{Quantifying the goodness of the one-dimensional system reduction. 
			The red point indicates $(x_{eff}, \beta_{eff})$ corresponding to the stationary state variables of Eq. (\ref{eq:GenDyn}) for a given network $\mathbf{A}$.
			The blue curve is the analytical stationary solution of the one-dimensional effective function Eq. (\ref{eq:GenDynEff}).
			The vertical and horizontal distances ($err_x$ and $err_{\beta}$) between the point and the curve represent the error of the analytical approximation.}\label{fig:Diagram}
	\end{center}
\end{figure}

For each realization of the stochastic interaction matrix, we can define two errors (see Figure \ref{fig:Diagram}) measuring the vertical and horizontal distance from the point ($x_{eff}$, $\beta_{eff}$) and the stationary solution of the one-dimensional resilience function $x(\beta)$. For the GLV dynamics, both errors become
\begin{equation}
\label{eq:ErrXDef}
    err = \frac{x_{eff} - x(\beta_{eff})}{x_{eff}}
        = \frac{\beta_{eff} - \beta(x_{eff})}{\beta_{eff}} 
        = 1 + \frac{\alpha}{x_{eff} \beta_{eff}}= 1-\frac{n}{d} 
\end{equation}
where $n = \sum_{ijkl} A_{ij}A_{kl}$, $d = S \cdot \sum_{ijk} A_{ij}A_{jk}$ and the $A_{ij}$ are the entries of the interaction matrix $\mathbf{A}$.

By taking $\mathbf{A}$  to be a random matrix, the error itself becomes a random variable whose 
probability distribution is inherited from the distribution of the random matrix. 
We can calculate the expected value and variance analytically under the assumption that the 
expected values of numerator and denominator in the terms above can be taken independently of each other.
After making this approximation, we get the expected value and variance of the error:
\begin{align}
    \label{eq:errExp}
    \left<err\right> &= \left<1 - \frac{n}{d}\right> \approx 1 - \frac{\left<n\right>}{\left<d\right>}\\
    \label{eq:errVar}
    \mathrm{Var}(err) &= \left<\left(\frac{d \left<n\right> - n \left<d\right>}{d \left<d\right>}\right)^2\right> \approx \frac{\left<d^2\right> \left<n\right>^2 - 2 \left<nd\right>\left<n\right> \left<d\right> + \left<n^2\right>\left<d\right>^2}{\left<d^2\right>\left<d\right>^2}
\end{align}


Hence, we need to calculate the terms $\left<d\right> = S \cdot \left< \sum_{ija} A_{ia}A_{aj} \right>$, $\left<n\right> = \left< \sum_{ijkl} A_{ij}A_{kl} \right>$, $\left<d^2\right> = S^2 \cdot \left< \sum_{ijklab} A_{ia}A_{aj}A_{kb}A_{bj} \right>$, $\left<n \cdot d\right> = S \cdot \left< \sum_{ijklmna} A_{ij}A_{kl}A_{ma}A_{an} \right>$ and $\left<n^2\right> = \left< \sum_{ijklmnop} A_{ij}A_{kl}A_{mn}A_{op} \right>$, where all indices are iterated over $\{1, 2, ..., S\}$.  
In full generality, we assume that all pairs of off-diagonal elements $(A_{i j}$ and $A_{j i})$ are drawn from a bivariate distribution with mean $\mu$, standard deviation $\sigma$ and correlation coefficient $\rho$. 
The diagonal elements are either drawn from a univariate distribution following the same statistics as the unconditional 
off-diagonal elements or kept fixed and constant by setting $A_{i i} = -d_i$. 
Under this setting one can generate both directed and undirected networks, being able to tune also the interaction properties \cite{suweis2014disentangling}. Then for the different cases we can 
quantitatively predict the errors of Gao et al. framework with respect to the actual quantities measured directly from the network.

\section{Discussion and Results}

\subsection{Stable GLV dynamics.}
We now discuss a subtle, but important issue related to the existence of a reachable stable point in the multi-dimensional GLV dynamics.
Indeed, depending on the parametrization of the adjacency matrix $\mathbf{A}$, Eq. (\ref{eq:GLVDyn}) may not have any stable stationary solutions. 
Recently, the width of this parameter region was also discussed in \cite{bunin2016interaction}.
However, we find that if we apply the multi-dimensional reduction to these unstable systems, we still find an effective one-dimensional equation with feasible and stable solutions.
In other words, the feasibility and stability of Eq. (\ref{eq:GLVSol}) does not imply that the corresponding solution of the full system given by Eq. (\ref{eq:GLVDyn}) is feasible and stable.
The map in this case is not well defined, as $x_{eff}$ can not be reach by the full dynamics.
Therefore, in order to have a meaningful multi-dimensional reduction, we must restrict our analysis only to those random matrices $\mathbf{A}$ that assure stability (and feasibility) of the complete GLV dynamics  (this issue is not discussed in \cite{gao2016universal}).

By combining our framework presented in section \ref{GLVsec} with results on the D-stability of random matrices \cite{allesina2012stability,tang2014reactivity,grilli2017feasibility}, we can achieve this goal.
If the off-diagonal elements of $\mathbf{A}$ are given by a distribution with mean $\mu$, standard deviation $\sigma$ and correlation coefficient $\rho$ and the diagonal elements are all fixed to a constant ($A_{ii} = -d$), we could set $d$ so that the analytic solution of the multi-dimensional GLV dynamics is stable and feasible (see Appendix \ref{app:StabilityCriteria}). 
For $\mu \leq 0$, the critical value to have stable GLV dynamics is $d_c = \sigma \sqrt{2S(1+\rho)} - \mu$ (that is $d$ of order $\sqrt{S}$ - rows 5-7 in Table \ref{tab:results}). 
For $\mu > 0$, stable GLV dynamics are assured if $d_c = (S-1)\mu$, that is $d$ of order $S$ (rows 8-10 in Table \ref{tab:results}). 

Note that for the case of a constant diagonal close to the critical value, shown in Figure \ref{fig:error_GLV_panel} (D), the theoretical value is not expected to give a good approximation to the empirical average, since in this case, the expected value of the denominator $\left<d\right>$ becomes zero.
In this case, the approximation of taking numerator and denominator separately is not justified.
Furthermore, sampling becomes difficult, as outliers may govern the empirical mean and standard deviation.

\subsection{Results for GLV dynamics with random interaction matrix}
The analytic derivation is complicated and tedious. Even in the simplest version of random matrix $A$, the entries $A_{ij}$ are all i.i.d., we need to separate out pairs $A_{ij}^2$ as they will lead to contributions other than $\mu^2$ where $\mu = \left<A_{ij}\right>$ (and similarly for higher order tuples). In order to do so, we devised an algorithm to solve it. The analytical expressions of expected value and variance of the error for different cases of interaction matrices $A$ at the highest order in the network size $S$ are listed in Table \ref{tab:results}.

\begin{table}[!htpb]
	\begin{tabular}{|p{5cm}|p{5cm}|p{5cm}|}
		\hline
		\textbf{Case} & $\left<err\right>$ & $\mathrm{Var}(err)$ \\
		\hline
		$A_{ij}$ i.i.d.                    & $0$ (exact)                     & $\frac{\sigma^4}{S^3 \mu^4}$ \\
		
		Correlation ($\rho=0$): &&\\
		$\rho = \mathrm{corr}(A_{ij}, A_{ji}) \in [-1, 1]$
		& $\frac{\rho \sigma^2}{S\mu^2}$
		& $\frac{\sigma^4}{S^3 \mu^4} \left(2\frac{\mu^2}{\sigma^2} \rho + (\rho - 1)^2\right) $ \\
		
		Constant diagonal: &&\\
		$A_{ii} = -d$ of order $1$ or $\sqrt{S}$
		& $\frac{\sigma^2\left((S - 2)\rho - 1\right)}{S^2\mu^2}$
		& $\frac{\sigma^4}{S^3 \mu^4} \left(2\frac{\mu^2}{\sigma^2} \rho + (\rho - 1)^2\right) $ \\
		$\qquad$ for $\rho = 0$
		& -$\frac{\sigma^2}{S^2\mu^2}$
		& $\frac{\sigma^4}{S^3 \mu^4} $ \\
		$\qquad$ for $\rho \ne 0$ and $S\gg1$
		& $\frac{\sigma^2\rho}{S\mu^2}$
		& $\frac{\sigma^4}{S^3 \mu^4} \left(2\frac{\mu^2}{\sigma^2} \rho + (\rho - 1)^2\right) $ \\
		
		$A_{ii} = -d$ of order $S$
		& $\frac{\sigma^2\left((S - 2)\rho - 1\right)}{S^2\mu^2\left(\frac{d}{d_c} - 1\right)^2}$
		& $\frac{\sigma^4}{S^3 \mu^4\left(\frac{d}{d_c} - 1\right)^4}\left(2\frac{\mu^2}{\sigma^2} \rho + (\rho - 1)^2\right)$ \\
		$\qquad$ for $\rho = 0$ and $S\gg1$
		& -$\frac{\sigma^2}{S^2\mu^2\left(\frac{d}{d_c} - 1\right)^2}$
		& $\frac{\sigma^4 }{S^3 \mu^4\left(\frac{d}{d_c} - 1\right)^4} $  \\
		$\qquad$ for $\rho \ne 0$ and $S\gg1$
		& $\frac{\sigma^2\rho}{S\mu^2\left(\frac{d}{d_c} - 1\right)^2}$
		& $\frac{\sigma^4}{S^3 \mu^4\left(\frac{d}{d_c} - 1\right)^4}\left(2\frac{\mu^2}{\sigma^2} \rho + (\rho - 1)^2\right)$ \\ \hline
	\end{tabular}
	\caption{Resulting analytical expressions, approximated to highest order in S}\label{tab:results}
\end{table}

The results in Table \ref{tab:results} can be summarized as follow: :
\begin{itemize}
	\item In all cases, the error (or its fluctuations) grows without bound if the ratio $\frac{\mu}{\sigma}$ goes to zero for a given network size $S$.
	\item The order of the fluctuations (namely $S^{-\frac{3}{2}}$) remains the same for all cases, while the order of the expected value changes. In particular, for interaction matrices $A$ without correlation ($\rho = 0$), the term dominating the error for large $S$ are the fluctuations while the mean value is either zero (for i.i.d. entries $A_{ij}$) or of order $S^{-2}$ (in case of a constant diagonal). On the other hand, for networks with non-zero correlation, the mean becomes the dominating term of order $S^{-1}$.
	\item  If the diagonal is of the same scale as $S$, the error may explode. This happens if $A_{ii} = - d_c$, where $d_c = (S - 1) \mu$  corresponds to the value of $d$ where the interaction matrix becomes stable and non-reactive for positive $\mu$.
\end{itemize}

We note that, differently from what is predicted by Gao et al., the approximation does not work for any positive interaction matrix $\mathbf{A}$.
In fact, on the one hand, our condition extends the validity of Gao et al. framework for matrix $\mathbf{A}$ with an asymmetric mixture of positive and negative interactions, as far as $\mu$ is not close to zero. 
Indeed, we can now understand that the stringent hypothesis on the positivity of the interactions assumed in Gao et al seminal work is not necessary.
At the same time our results highlight that if matrix $\mathbf{A}$ has a very large variance with respect to $\mu$ and $S$ is not large enough, then the collapse will fail.
For example, if interactions strengths are very heterogenous (e.g. power law distributed), although mutualistic (positive), the system resilience can not be described by the one-dimensional analytical resilience function.

\begin{figure}[!htpb]
    \centering
    \includegraphics[width=1\linewidth]{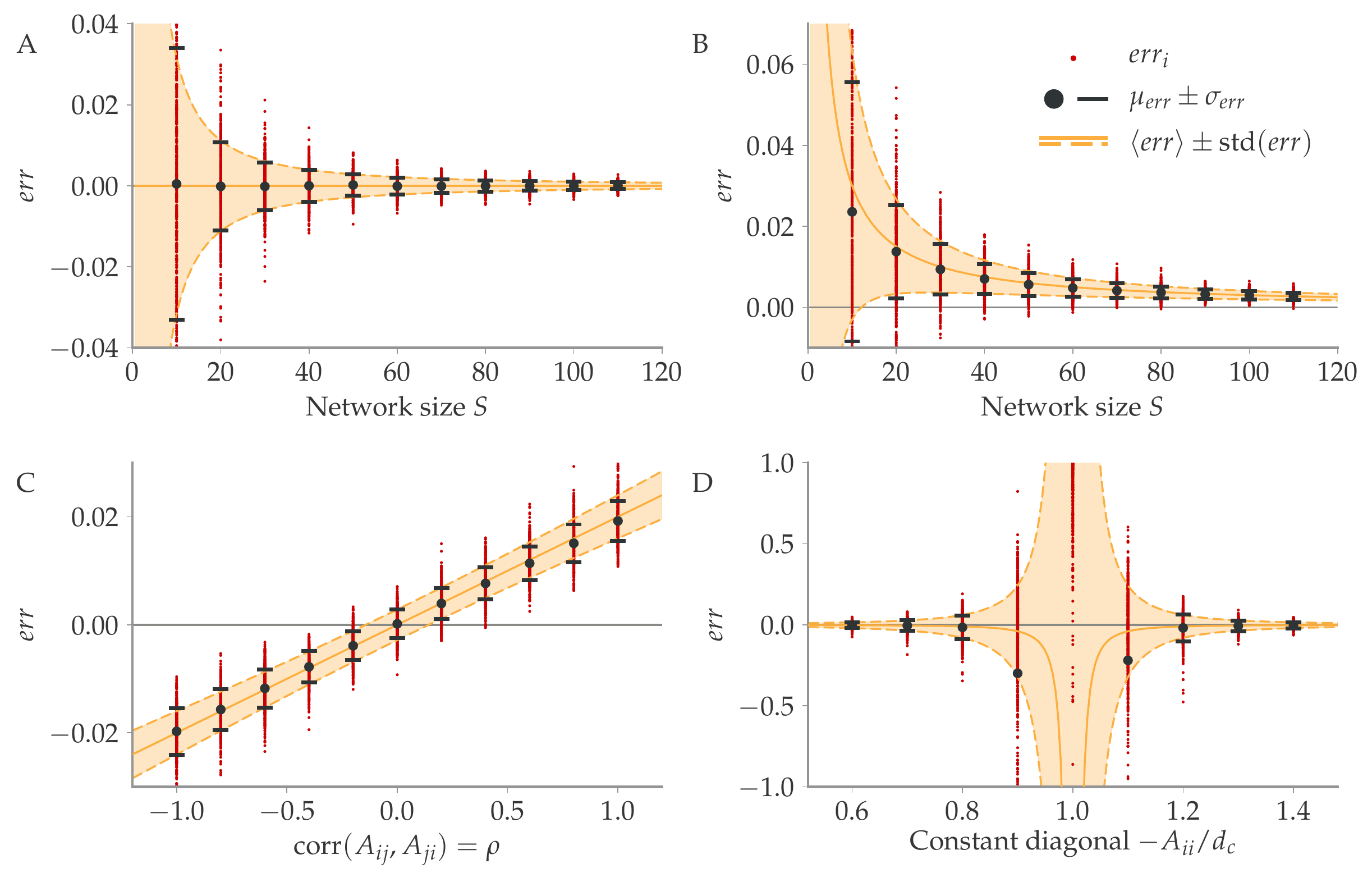}
    \caption{
        Comparison of theoretical results with numerical samples.
        Each red dot corresponds to the error calculated for one interaction matrix. 
        We sampled 500 matrices independently and calculated empirical means $\mu_{err}$ 
        and standard deviations $\sigma_{err}$ (plotted as black dots and bars, respectively).
        The theoretical mean $\left<err\right>$ is plotted as a orange line, the shaded area indicates the 
        predicted standard deviation $\mathrm{std}(err)$.
        For all figures, the entries of $\mathbf{A}$ are drawn from a normal distribution with
        $\mu = \sigma = 1$. The upper plots show the effect network size on the error 
        in the case of (A) all elements drawn i.i.d. or (B) with positive 
        correlation $\rho = 0.3$. 
        The lower plots show the error for networks of size $S = 50$ for 
        (C) varying correlation $\rho$ or (D) enforcing a constant diagonal $A_{ii}$
        relative to the critical value $d_c$.
    }
    \label{fig:error_GLV_panel}
\end{figure}

In order to test these analytical results, we sampled the interaction matrix with the corresponding statistics numerically and compared the empirical mean and standard deviation with the theoretical predictions. 
The results can be observed in Figure \ref{fig:error_GLV_panel}. 
In all cases, the theoretical predictions are met very well.
There is a notable but small deviation for small network sizes  $S = 20$, namely slight underestimation of the mean for the case of correlation, c.f. plot B of Figure \ref{fig:error_GLV_panel}.

In our discussion, we set the connectivity (the fraction of non-zero elements) to one, i.e. $C=1$. 
Generalizing our results to not fully connected networks is straightforward. We model sparsely connected
networks by drawing a mask $\mathbf{T}$ with entries drawn from a Bernoulli distribution, $T_{ij} \sim \mathcal{B}(1, C)$, 
independently drawing another matrix $\mathbf{A}'$ with specific statistics as before, and 
finally setting $\mathbf{A} = \mathbf{T} \circ \mathbf{A}'$, where $\circ$ denotes the Hadamard or entry-wise product. 
Since $A_{ij}$ and $T_{ij}$ are independent, it suffices to insert the moments $\left< T_{ij}^k \right>$, into the 
calculation of $err$ and its variance, c.f. Eqs. \ref{eq:errExp} and \ref{eq:errVar}. For the case of correlated
pairs discussed above, the expression of the expected value remains the same, $\left< err \right> = \frac{\rho \sigma^2}{S \mu^2}$, 
while the variance is increased, 
\begin{equation}
    \label{eq:var_sparse}
    \mathrm{Var}(err) = \frac{\sigma^4}{S^3 \mu^4} \left(2\frac{\mu^2}{\sigma^2} \left(1 - C + C\rho\right) + (1 - C \rho)^2 + \left(1 - C \right)^2 \frac{\mu^4}{\sigma^4} \right) \,.
\end{equation}
This is to be expected, as for non-zero mean the sparse mask $\mathbf{T}$ contributes to the variance. 

Finally, our results are robust for other definitions of error. In the Appendix \ref{app:AnotherDef}, we also provide the analytical expressions of another error definition, i.e. the distance from the mean point $\overline{err} = err(\left<x_{eff}\right>,\left<\beta_{eff}\right>$).

\subsection{Beyond GLV dynamics}
In the most general setting, the stationary solution of Eq. (\ref{eq:GenDynEff}) is $\beta(x) = -\frac{F(x)}{G(x, x)}$. 
If we use the error definition $\overline{err}$ (see  Appendix \ref{app:AnotherDef} for more details), we can have some qualitative insights on the conditions under which the multi-dimensional collapse is expected to work also for more general dynamics than the GLV discussed above. 

In fact, for the general dynamics given by Eq. (\ref{eq:GenDyn}), then the following equation holds:
\begin{equation}
\overline{err} = \left|1 + \frac{\left<F(x_{eff})\right>}{\left<\beta_{eff}\right> \left<G(x_{eff}, x_{eff})\right>}\right|
\end{equation}

For GLV dynamics, the key quantity in determining the feasibility of the multi-dimensional reduction is a simple function of the product between $\rho$ and $\sigma/\mu$ compared to the system size $S$ \cite{pigolotti2013species}.
We thus ansatz the possibility that this quantity is crucial in determining the quality of the collapse also for different type of dynamics.

If the random matrix $\mathbf{A}$ is generated by i.i.d. random variables ($A_{ij}=p(\mu,\sigma)$) and Eq. \ref{eq:condForAnotherDef} $S >> \frac{\sigma}{|\mu|}$ holds, then we find through Eq. (\ref{eq:betaEffDef}) that $\left<\beta_{eff}\right> \approx \frac{S^2 \mu^2 + \sigma^2}{S \mu} \approx S\mu$ does not depend on the specific dynamics (see Appendix C).

Therefore, we obtain the following equation:
\begin{equation}\label{eq:ErrBetaGen}
err_{\beta} = \left|1 + \frac{\left<F(x_{eff})\right>}{S \mu \left<G(x_{eff}, x_{eff})\right>}\right|.
\end{equation}
We note that Eq. (\ref{eq:ErrBetaGen}) goes to zero, clearly depending on the functions $F(x_{eff})$ and $G(x_{eff},x_{eff})$.
In other words, the results presented by Gao et al. hold only for particular choices of $F(x_i)$ and $G(x_i,x_j)$, i.e. those for which $\frac{\left<F(x_{eff})\right>}{S \mu \left<G(x_{eff},x_{eff})\right>}\approx -1$.
In brief, for general dynamics, if the $S >> \frac{\sigma}{|\mu|}$ does not hold, the collapse will fail (e.g. GLV dynamics); if it holds and $\frac{\left<F(x_{eff})\right>}{S \mu \left<G(x_{eff},x_{eff})\right>}\approx -1$, the collapse will work.

\begin{figure}[!htpb]
	\subfigure[\textit{Random case} $\mathbf{A}$ is drawn from a normal distribution with $\mu=0.2$ and $\sigma=1$.]{\label{fig:error_Gen_Normal}
		\includegraphics[width=0.47\columnwidth]{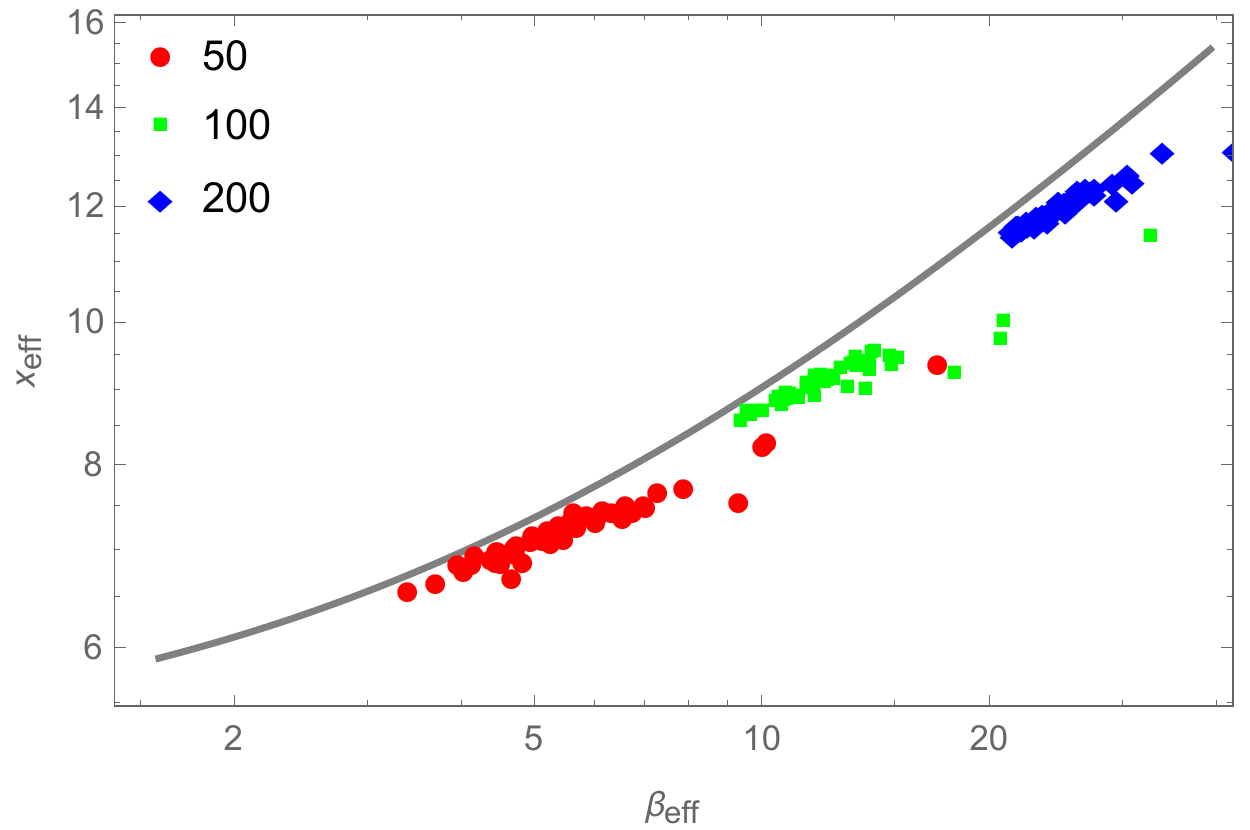}
	}    
	\subfigure[\textit{Mutualistic case} $\mathbf{A}$ is drawn from a lognormal distribution with $\mu=0.096$ and $\sigma=4.253$.]{\label{fig:error_Gen_LogNormal}
		\includegraphics[width=0.47\columnwidth]{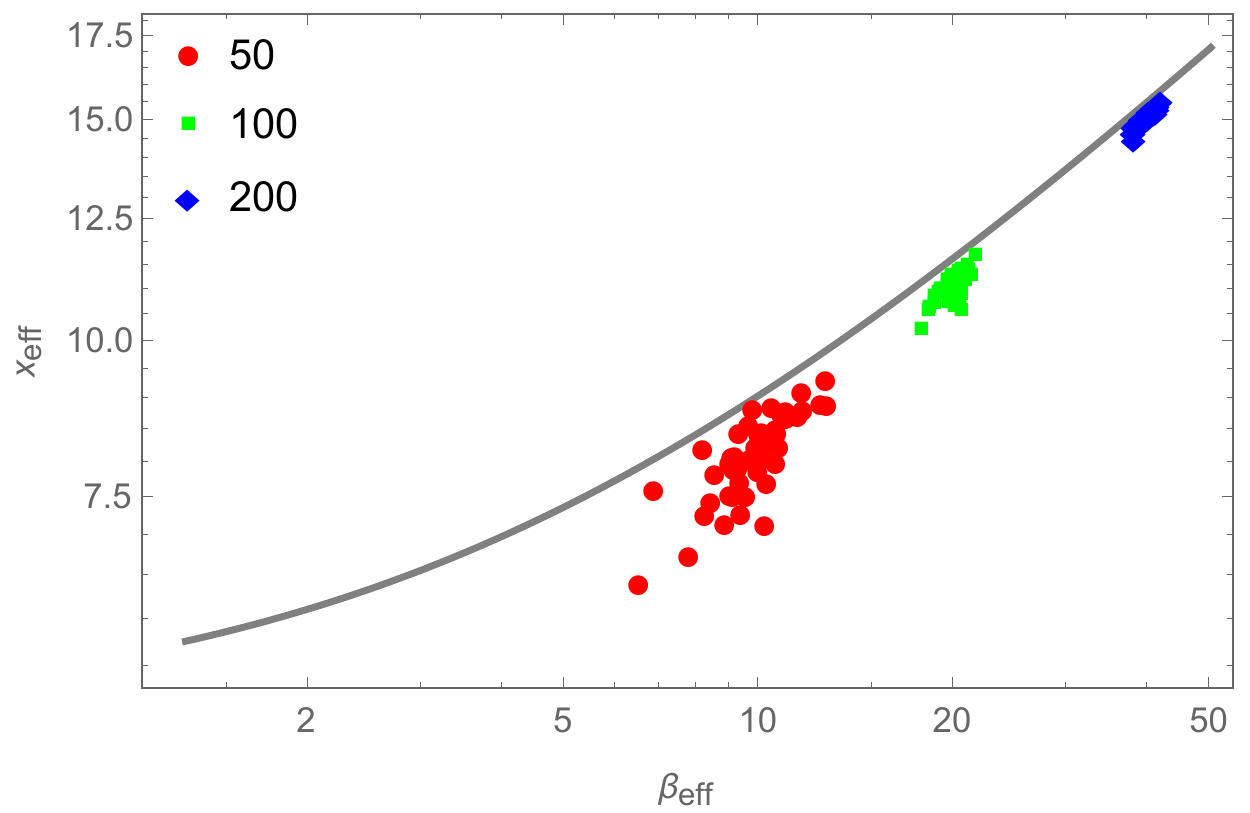}    
	}
		
	\caption{Collapse of the S-dimensional system equations for the non-linear dynamics given by Eq. (\ref{Gaodyn}).
		Random networks of different sizes 50 (red circle), 100 (green square), 200 (blue diamond).
		We observe that, as predicted, when the heterogeneity in the interaction strength is very high ($\sigma\gg\mu$), the collapse fails.
	}\label{fig:error_Gen}
\end{figure}

In Figure \ref{fig:error_Gen}, we test the above results by using the dynamics for ecological communities proposed in \citet{gao2016universal}:
\begin{equation}\label{Gaodyn}
\frac{dx_i}{dt} = B + x_i \left( 1-\frac{x_i}{K} \right) \left( \frac{x_i}{C}-1\right) +\sum_{j=1}^S A_{ij}\frac{x_i x_j}{D+E x_i+H x_j},
\end{equation}
where $B=0.1, C=1, K=5, D=5, E=0.9, H=0.1$. We show that the collapse may work also for both positive-negative interactions $A_{ij}$ if $S$ is large enough shown in Figure \ref{fig:error_Gen_Normal}. 
On the other hand, Figure \ref{fig:error_Gen_LogNormal} confirms that condition $(i)$ and $(ii)$ and the positivity of the interactions of $\mathbf{A}$ are not sufficient to guarantee the validity of the one-dimensional approximation also for dynamics beyond GLV: If the matrix $\mathbf{A}$ has a very large variance, the collapse fails also for the specific dynamics used by Gao et al..

\section{Conclusions}

In this paper we have shown under which condition a large dynamical system can be effectively approximated with one-dimensional equation.
The order parameter that appears as a variable in the effective equation can be obtained from a simple expression of the local variables.
Under this approximation, it becomes clear which properties of the interactions determine the state of the system and it turns out to be possible to quantify their effects.

We explored which properties of the interactions determine the accuracy of the approximation.
In general, the form of the multi-dimensional equations and how their non-linearities are introduced will influence the opportunity to approximate the original set of equations with the corresponding one-dimensional equation.
In order to focus on the effect of the interactions, we therefore first have considered a simple idealized scenario -- the generalized Lotka-Volterra equations -- where the interactions are linear. 
In this context, the accuracy of the approximation is only determined by the interaction matrix.

The criterion we obtained relates the variability of the interactions between the agents/nodes and the number of the agents.
In particular, for the approximation to work, the size of the system has to be larger than a critical value proportional to the coefficient of variation of the interaction strengths.
Also the reciprocity of interactions plays an important role: the approximation is expected to work for any interaction strengths if there is not any correlation in the activity between each pair of nodes in the network.
As the correlation between reciprocal interactions is increased, the larger the size of the system must be so to guarantee the accuracy of the approximation.

Finally we have shown that the approximation works also for interaction matrices with a mixture of positive and negative signs and that it can be extended to more complicated and non-linear dynamics. These results open up possible applications of the framework to food-webs, neuronal networks and social/economic interactions.

\clearpage

\appendix

\section{One-dimensional effective equation}\label{app:EffFunDer}

We here summarize the mathematical details of the one dimensional reduction proposed by Gao et al. \cite{gao2016universal}.

For the class of dynamics described by Eq. (\ref{eq:GenDyn}), we first consider a scalar quantity $y_j$.
A neighbour $j$ is selected with probability proportional to the outgoing degree of $j$ $s_j^{out} = \sum_{i=1}^{S} A_{ij}$ and the mean over all nearest neighbour nodes is $\left<y_j\right>_{nn} = \frac{ \frac{1}{S} \sum_{j=1}^S s_j^{out} y_j}{\frac{1}{S} \sum _{j=1}^S s_j^{out}}$.
Selecting $y_j(x_i) = G(x_i, x_j)$, we could write the second term of the right part of Eq. (\ref{eq:GenDyn}) as following: $\sum_{j=1}^S A_{ij}G(x_i,x_j) = s_i^{in}\left<y_j(x_i)\right>_{j~nn~of~i}$, where $s_i^{in} = \sum_j^S A_{ij}$ is the ingoing degree.
If the degree correlations of the network described by $A$ are small, then the neighborhood of $i$ is on average identical to the neighborhood of all other nodes and the relation $s_i^{in}\left<y_j(x_i)\right>_{j~nn~of~i} = s_i^{in}\left<y_j(x_i)\right>_{nn}$ holds for each $i$ and $j$.
To formalize the above analysis the operator $\mathcal{L}(y) = \mathbf{\frac{1^T A y}{1^T A 1}} = \frac{ \frac{1}{S} \sum_{j=1}^S s_j^{out} y_j}{\frac{1}{S} \sum _{j=1}^S s_j^{out}}$ can be introduced, where $\mathbf{1} = (1,...,1)^T$ is the unit vector.
According to this operator, Eq. (\ref{eq:GenDyn}) can be written as $\frac{dx_i}{dt} = F(x_i) + s_i^{in} \mathcal{L}\left(G(x_i,\mathbf{x})\right)$.
If $G(x_i, x_j)$ is linear in $x_j$ or the variance in the components of $\mathbf{x}$ is small, then $\mathcal{L}\left(G(x_i, \mathbf{x})\right) \approx G(x_i, \mathcal{L}(\mathbf{x}))$.
Therefore $\frac{dx_i}{dt} = F(x_i) + s_i^{in} G\left(x_i, \mathcal{L}(\mathbf{x})\right)$ or, in vector notation, $\frac{d\mathbf{x}}{dt} = F(\mathbf{x}) + \mathbf{s}^{in} \circ G\left(\mathbf{x}, \mathcal{L}(\mathbf{x})\right)$.
By applying the operator to both sides of the latter equation we have: $\frac{d\mathcal{L}(\mathbf{x})}{dt} = \mathcal{L}\left(F(\mathbf{x}) + \mathbf{s}^{in} \circ G(\mathbf{x}, \mathcal{L}(\mathbf{x}))\right) \approx F\left(\mathcal{L}(\mathbf{x})\right) + \mathcal{L}(\mathbf{s}^{in})G\left( \mathcal{L}(\mathbf{x}), \mathcal{L}(\mathbf{x}) \right)$.
At last, we obtain the one-dimensional effective equation $\dot{x} = F(x) + \beta G(x, x)=f(x,\beta)$.
By solving the equilibrium state of this equation ($f(x,\beta)= 0$), we could obtain the resilience curve $x(\beta)$ or $\beta(x)$ in the two dimensional coordinate system.
We then calculate directly $x_{eff} = \mathbf{\frac{1^{T} A x}{1^{T} A 1}} = \frac{\sum_{ij} A_{ij} x_j}{\sum_{ij} A_{ij}}$ and $\beta_{eff} = \mathbf{\frac{1^{T} A s^{in}}{1^{T} A 1}} = \frac{\sum_{ij} A_{ij} A_{ji}}{\sum_{ij} A_{ij}}$ through the interaction matrix $A$ of the original multi-dimensional dynamics.
If the point $(x_{eff},\beta_{eff})$ lies on the resilience curve, then the collapse works; If not, it fails.
Figure \ref{fig:Diagram} is a diagram illustrating how the goodness of the one-dimensional approximation can be quantified by $err_x$ and $err_\beta$, i.e. the distance of the point $(x_{eff},\beta_{eff})$ to the resilience curve $x(\beta)$.

\section{Stability criteria for random matrices}\label{app:StabilityCriteria}

As shown in \cite{grilli2017feasibility}, a feasible fixed point $\mathbf{x}^{*}$ of the GLV dynamics
(i.e. one with all entries $\mathbf{x}^*_i \ge 0$)  is globally 
stable if the symmetrized interaction matrix $\mathbf{A} + \mathbf{A^T}$ is negative definite. 
A sufficient condition for this negative definiteness in case of random matrices used in this 
study is derived in \cite{tang2014reactivity}: It can be achieved by setting the diagonal elements
to a constant value $A_{ij} = -d$, where $d$ has to be larger than some critical value $d_c$. 
In terms of the mean $\mu$, variance $\sigma^2$ and correlation coefficient $\rho$, this critical 
value is found to be 
\begin{equation}
    d_c = 
    \begin{cases}
        (S-1)\mu & \text{if} ~ \mu>0 ;\\ 
        \sigma \sqrt{2S(1+\rho)} - \mu& \text{if} ~ \mu \leq 0 .
    \end{cases}
\end{equation}

\section{Error as distance from the mean point}\label{app:AnotherDef}

Now we provide the analytical expression of another error definition according to the mean point. 
Indeed, we can define the error as the distance from the mean point $(\left<x_{eff}\right>,\left<\beta_{eff}\right>)$ to the stationary solution of the one-dimensional resilience function $x(\beta)$ as following: $\Vert  (\left<x_{eff}\right>,\left<\beta_{eff}\right>), x(\beta) \Vert$, where $\left<x_{eff}\right>$ and $\left<\beta_{eff}\right>$ are the mean of several realizations of $x_{eff}$ and $\beta_{eff}$ calculated from Eqs. (\ref{eq:xEffDef})-(\ref{eq:betaEffDef}).
The vertical and horizontal distance from the mean point $( \left<x_{eff}\right>, \left<\beta_{eff}\right> )$ to the resilience function $x(\beta)$ is $err_x = \left|\frac{\left<x_{eff}\right> - x\left(\left<\beta_{eff}\right>\right)}{\left<x_{eff}\right>}\right|$ and $err_{\beta} = \left|\frac{\left<\beta_{eff}\right> - \beta\left(\left<x_{eff}\right>\right)}{\left<\beta_{eff}\right>}\right|$. 
For GLV dynamics given by Eq. (\ref{eq:GLVDyn}), the resilience function is Eq. (\ref{eq:GLVSol}). 
Therefore $err_x = err_{\beta} = \left|1 + \frac{\alpha}{ \left<x_{eff}\right> \left<\beta_{eff}\right> }\right|$. 
Our results discussed in main text are also robust for this error definition.

\textbf{Off-diagonal drawn from a bivariate distribution.} If all pairs of off-diagonal elements $(A_{i j}$ and $A_{j i})$ are drawn from a bivariate distribution with mean $\mu$, standard deviation $\sigma$ and correlation coefficient $\rho$, and diagonal elements $A_{i i} = -d_i$ are kept fixed.
We will use the following approximate equations which would strictly hold only in the very large $S$: $\mu = \frac{1}{S(S-1)} \sum_{i\neq j} A_{i j}$, $\sigma^2 = \frac{1}{S(S-1)} \sum_{i\neq j} A_{i j}^2 - \mu^2$, $\rho \sigma^2 = \frac{1}{S(S-1)} \sum_{i\neq j} A_{i j}A_{j i} - \mu^2$ where S is the matrix size.
Then we could get the following approximate equations: $\sum_{i j} A_{i j} = \sum_i A_{i i} + \sum_{i\neq j} A_{i j} = \sum _i d_i + S(S-1)\mu$ and $\sum_{i j k} A_{i k} A_{k j} = \sum_i (-d_i)^2 + (S-1)[2 \mu  \left(\sum _i -d_i\right) + S(S-1)\mu^2 + S \rho \sigma^2]$. 

For GLV dynamics the analytical solution for the equilibrium state is $\mathbf{x^{*} = -A^{-1} \cdot \boldsymbol{\alpha}}$ where $\boldsymbol{\alpha}$ is a vector whose components are all equal to the constant $\alpha$, so $\sum_{i j} A_{i j} x_j = -S \alpha$.
According to the definition $x_{eff} = \frac{\sum_{i j} A_{i j} x_j}{\sum_{i j} A_{i j}}$ and $\beta_{eff} = \frac{\sum_{i j k} A_{i k} A_{k j}}{\sum_{i j} A_{i j}}$, we could get following equations: $\left<x_{eff}\right> = \frac{-S \alpha}{\sum_i (-d_i) + S(S-1)\mu}$ and $\left<\beta_{eff}\right> = \frac{\sum_i (-d_i)^2 + (S-1)[2 \mu \sum_i (-d_i) + S(S-1)\mu^2 + S \rho \sigma^2]}{\sum_i (-d_i)+S(S-1)\mu}$.

\textbf{Off-diagonal drawn from a bivariate distribution and diagonal elements set to a constant.}
If the diagonal elements of $\mathbf{A}$ are the same constant ($A_{ii} = -d$), then $\left<x_{eff}\right> = \frac{-\alpha}{(-d) + (S-1)\mu}$ and $\left<\beta_{eff}\right> = \frac{(-d)^2 + (S-1)[2 \mu (-d) + (S-1)\mu^2 + \rho \sigma^2]}{(-d) + (S-1)\mu}$. 

\textbf{Off-diagonal drawn from a bivariate distribution and diagonal elements drawn from a univariate distribution.} If the diagonal elements $A_{ii} = -d_i$ are i.i.d. random variables with given distribution of mean $\mu_d$ and standard deviation $\sigma_d$, then $\left<x_{eff}\right> = \frac{-\alpha}{\mu_d + (S-1)\mu}$ and $\left<\beta_{eff}\right> = \frac{(\mu_d)^2+(\sigma_d)^2 + (S-1)[2 \mu \mu_d + (S-1)\mu^2 + \rho \sigma^2]}{\mu_d + (S-1)\mu}$. 

\textbf{i.i.d. independent random variables.}
If the random matrix $\mathbf{A}$ is generated by i.i.d. random variable ($A_{ij}=p(\mu,\sigma)$), then the distribution of diagonal is the same as non-diagonal ($\mu = \mu_d$ and $\sigma = \sigma_d$).
Therefore we have $\left<x_{eff}\right> = \frac{-\alpha}{\mu_d + (S-1)\mu} = \frac{-\alpha}{S \mu}$ and $\left<\beta_{eff}\right> = \frac{(\mu_d)^2+(\sigma_d)^2 + (S-1)[2 \mu \mu_d + (S-1)\mu^2 + \rho \sigma^2]}{\mu_d + (S-1)\mu} = \frac{\mu^2 + \sigma^2 + (S-1)[(S+1)\mu^2 + \rho \sigma^2]}{S \mu} \approx \frac{S^2 \mu^2 + \sigma^2}{S \mu}$. 
Finally, $err_x = err_{\beta} = \left|1 + \frac{\alpha}{ \left<x_{eff}\right> \left<\beta_{eff}\right> }\right| = \frac{\sigma^2}{S^2 \mu^2+ \sigma^2}$. 
If the following condition holds
\begin{equation}\label{eq:condForAnotherDef}
S >> \frac{\sigma}{|\mu|} 
\end{equation}
the collapse will work (i.e. $\left<err_x\right> = \left<err_{\beta}\right> \approx 0$); Otherwise, the collapse will fail.

\end{document}